\documentclass[prl,twocolumn,superscriptaddress,showpacs,floats,floatfix]{revtex4}

\usepackage{epsfig}
\begin{document}

\draft

\title{Three reversible states controlled on a gold monoatomic contact by the 
electrochemical potential}

\author{Manabu Kiguchi$^{1,2}$, Tatsuya Konishi$^1$, Kouta Hasegawa$^1$, Satoshi Shidara$^1$, and Kei Murakoshi$^1$}

\address{$^1$Division of Chemistry, Graduate School of Science, Hokkaido University, Sapporo, 060-0810, Japan}
\address{$^2$Precursory Research for Embryonic Science and Technology (PRESTO), 
Japan Science and Technology Agency, Sapporo, Hokkaido 060-0810, Japan}

\date{\today}

\begin{abstract}
Conductance of an Au mono atomic contact was investigated under the electrochemical potential 
control. The Au contact showed three different behaviors depending on the potential: 1 $G_{0}$  ($G_{0}$  = 
$2e^{2}/h$), 0.5 $G_{0}$ and not-well defined values below 1 $G_{0}$  were shown when the potential of the contact 
was kept at -0.6 V (double layer potential), -1.0 V (hydrogen evolution potential), and 0.8 V (oxide 
formation potential) versus Ag/AgCl in 0.1 M Na$_{2}$SO$_{4}$ solution, respectively. These three reversible 
states and their respective conductances could be fully controlled by the electrochemical potential. 
These changes in the conductance values are discussed based on the proposed structure models of 
hydrogen adsorbed and oxygen incorporated on an Au mono atomic contact.

\end{abstract}

\maketitle

\section{INTRODUCTION}
\label{sec1}

Study of charge transport in atomic scale metal or molecular nano wires is of fundamental 
interest, with potential applications for ultra-small electronic devices \cite{1,2}. In order to use these 
nano wires for devices, their conductance should be tunable by some external field, such as light or 
magnetic field. There are some experimental challenges to controlling the conductance of nano 
wires by an external field \cite{3,4,5,6,7}. The switching of a photochromic molecular wire from the 
conducting state to the insulating state was observed by radiation of visible light \cite{3}. A change in 
the conductance behavior by molecular adsorption was observed for Au, Pt, Cu, Fe, Co, and Ni 
nano wires \cite{4,5,6,7}. Here, it should be noted that the conductance can be switched from the off state to 
the on state or vice versa (among two states) in most studies. If the conductance of the nano wires 
could be controlled among three different conductance states, it would have attracted wide attention 
for fundamental science and technological applications.

In the present study, we paid attention to the electrochemical potential as an external field 
\cite{8,9,10-1,10-2,10}. 
The electrochemical potential can control the procession of two individual electrochemical 
reactions on a metal surface by changing the electrochemical potential of the metal electrode by 
only several volts. For example, hydrogen or oxygen evolution proceeds when the electrochemical 
potential of Pt electrodes is maintained at more negative than -0.2 V or more positive than +1.3 V 
versus Ag/AgCl in 0.1 M H$_{2}$SO$_{4}$ solution, respectively \cite{11}. The clean hydrogen or oxygen 
adsorbed states can be selectively prepared by controlling the electrochemical potential. Several 
states of the surface condition, at least three, can be maintained under the electrochemical potential 
control. If the conductance of the metal nano wire can be defined by the surface condition of the 
wire, the conductance of the metal nano wire may be controlled by the electrochemical potential. 
Au was investigated in the present study for the following reasons. First, the hydrogen and oxygen 
evolution can proceed without dissolving the Au ions from the Au electrode; this is a characteristic 
of the Au electrode in solution. In the case of Ni, Ni rapidly dissolves at the oxygen evolution 
potential \cite{11}. It is thus difficult to prepare stable Ni nano wires under oxygen evolution. Second, 
the Au mono atomic contact shows a well defined fixed conductance value. It normally shows 1 $G_{0}$ 
($2e^{2}/h$), and it shows 0.5 $G_{0}$ under hydrogen evolution \cite{8,10}. Showing well defined conductance 
values is useful for the understanding and utilization of the metal nano wire. Under ultra high 
vacuum (UHV) condition, the Au mono atomic contact does not show a well-defined fractional 
conductance value by the introduction of hydrogen molecule \cite{4}. In solution, the Au mono atomic 
contact shows well defined fractional conductance. Hydrogen incorporated wire, dimerized wire, 
and other atomic and electronic structures of the wire, were proposed to be the origin of the 
fractional conductance value \cite{8,10}. However, the actual origin is not clear at present. In addition, 
the effect of oxygen on the Au mono atomic contact, such as the structure, stability, and 
conductance, in solution is not clear. In the present study, we have studied the Au mono atomic 
contact under the electrochemical potential control to reveal the effect of hydrogen and oxygen on 
the conductance behavior and to control the conductance of the Au mono atomic contact among 
three different states.

\section{EXPERIMENTAL}
\label{sec2}

\begin{figure}
\begin{center}
\leavevmode\epsfysize=80mm \epsfbox{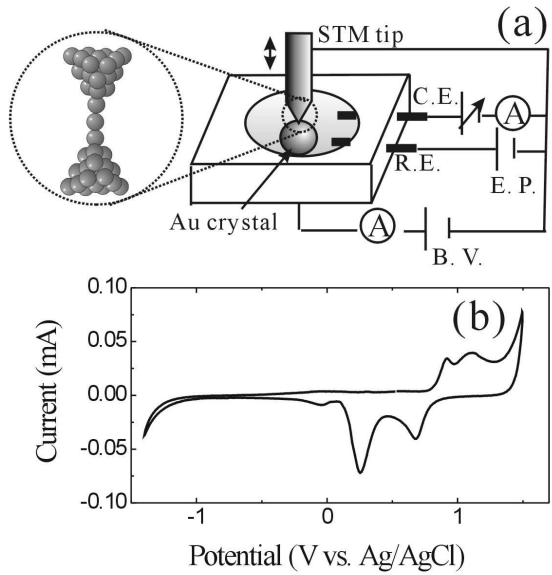}
\caption{
(a) Schematic view of the electrochemical STM, C.E.: counter electrode, R.E.: reference 
electrode, E.P.: electrochemical potential, B.V. Bias voltage, (b) Cyclic voltammogram of the Au 
electrode in 0.1M Na$_{2}$SO$_{4}$. } \label{fig1}
\end{center}
\end{figure}

The experiments were performed with the modified scanning tunneling microscope (STM: 
Pico-SPM, Molecular Imaging Co.) with a Nano ScopeIIIa controller (Digital Instruments Co.) in 
an electrochemical cell (see Ref \cite{10-1,10-2,10} for a detailed description of the experimental setup and 
conditions). The STM tip was made of an Au wire (diameter 0.25 mm, $>$ 99 $\%$) coated with wax to 
eliminate ionic conduction. The Au(111) substrate was prepared by a flame annealing and 
quenching method. Figure~\ref{fig1}(a) shows a schematic view of the experimental setup. The 
electrochemical potential ($\phi$) of the Au tip and substrate was controlled using a potentiostat 
(Pico-Stat, Molecular Imaging Co.) with an Ag/AgCl reference electrode. A 0.50 mm diameter Pt 
wire was used as a counter electrode. The STM tip was repeatedly moved into and out of contact 
with the Au substrate at a rate of 50 nm/s in an electrolyte solution. The electrolyte solution was 0.1 
M Na$_{2}$SO$_{4}$ or 50 mM H$_{2}$SO$_{4}$. Conductance was measured during the breaking process under an 
applied bias of 20 mV between the tip and substrate. All statistical data was obtained from a large 
number (over 1000) of individual conductance traces. Figure~\ref{fig1}(b) shows the cyclic voltammogram 
(CV) of an Au electrode in the 0.1 M Na$_{2}$SO$_{4}$   solution. The double layer regime extended from $\phi$ = 
-0.7 V to 0.8 V. When the electrochemical potential of the Au electrode was kept at a potential more 
positive than $\phi$ = 1.3 V, the oxygen evolution proceeded via the formation of an oxide layer on the 
Au electrode. The hydrogen evolution proceeded at a potential more negative than $\phi$ = -0.7 V 
\cite{11,12,13}. Conductance of the Au nano contacts was measured at $\phi$ = -0.6V (double layer potential), 
$\phi$ = -1.0 V (hydrogen evolution potential) and $\phi$ = 0.8 V (oxide formation potential).

\section{RESULTS AND DISCUSSION}
\label{sec3}

\begin{figure}
\begin{center}
\leavevmode\epsfysize=80mm \epsfbox{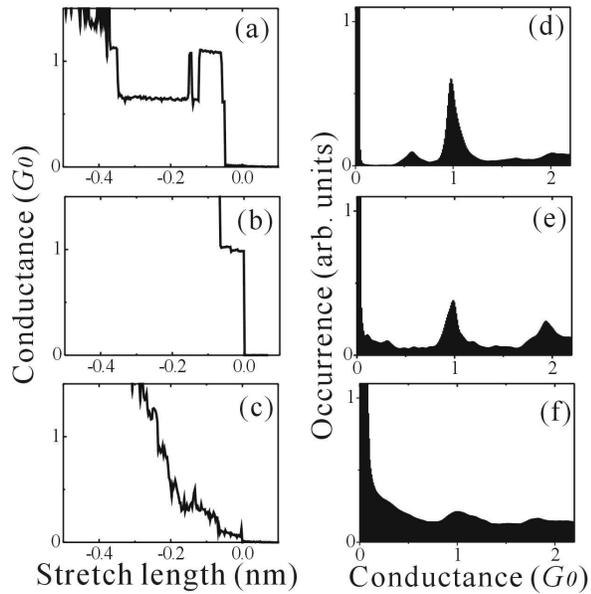}
\caption{
Conductance traces and histograms of the Au nano contacts in 0.1M Na$_{2}$SO$_{4}$ solution. The 
electrochemical potential was (a, d) $\phi$= -1.0 V (hydrogen evolution potential) (b, e) $\phi$= -0.6 V 
(double layer potential), and (c, f) $\phi$= 0.8 V (oxide formation potential). The conductance 
histograms were obtained from 5000 conductance traces of breaking the Au contacts. The intensity 
of the conductance histograms was normalized with the number of the conductance traces. 
} \label{fig2}
\end{center}
\end{figure}

Figure~\ref{fig2} shows the typical conductance traces and histograms of the Au nano contacts in 0.1 M 
Na$_{2}$SO$_{4}$ solution during breaking the contact at the double layer, hydrogen evolution and oxide 
formation potentials. At the double layer potential (Fig.~\ref{fig2}(b), (e)), the conductance decreased in a 
stepwise fashion with each step occurring at integer multiples of $G_{0}$. The corresponding 
conductance histograms showed a well-defined peak near 1 $G_{0}$, which corresponded to a clean Au 
atomic contact or wire \cite{2}. At the hydrogen evolution potential (Fig.~\ref{fig2}(a), (d)), reversible transition 
of the conductance between 1 $G_{0}$ and 0.5 $G_{0}$ was observed. This conductance fluctuation resulted in 
the 0.5 $G_{0}$ peak in the conductance histogram. At the oxide formation potential (Fig.~\ref{fig2}(c), (f)), the 
conductance continuously decreased stepwise at various conductance values below 1 $G_{0}$. The 
corresponding histogram showed a broad feature at 1 $G_{0}$ together with a large background. The 
conductance of the Au mono atomic contact could be controlled among three different conductance 
states (1 $G_{0}$, 0.5 $G_{0}$ and conductance below 1 $G_{0}$) by the electrochemical potential of the contact. 
These electrochemical potential dependent changes were fully reversible. The potential in which the 
0.5 $G_{0}$ peak (or large background) appeared in the conductance histogram agreed with the onset 
potential of the hydrogen evolution (or oxide formation) reaction on the Au electrode. In the 
conductance histogram, the intensity of the 0.5 $G_{0}$ peak normalized by the 1 $G_{0}$ peak increased, and 
then saturated, as the potential was scanned from the double layer potential to the hydrogen 
evolution potential. The intensity of the background continuously increased, as the potential was 
scanned from the double layer potential to the oxide formation potential. The 1 $G_{0}$ peak was not 
apparent in the conductance histogram at more positive than $\phi$ = 1.0 V.

\begin{figure}
\begin{center}
\leavevmode\epsfysize=60mm \epsfbox{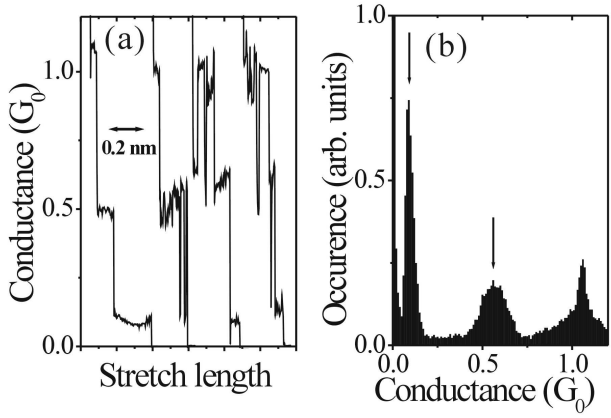}
\caption{
(a) Conductance trace and (b) histogram of the Au nano contacts in 50 mM H$_{2}$SO$_{4}$ solution. 
The electrochemical potential was $\phi$= -0.3 V (hydrogen evolution potential). The conductance 
histogram was obtained from 3000 conductance traces of breaking the Au contacts.} \label{fig3}
\end{center}
\end{figure}

The conductance behavior of the Au mono atomic contact was investigated in other 
electrolytes, 0.1 M NaCl, 0.1 M NaOH, 50 mM H$_{2}$SO$_{4}$, and 0.1 M HClO$_{4}$. In the conductance 
histogram, a large background below 1 $G_{0}$ was observed at the oxide formation potential, while the 
0.5 $G_{0}$ peak was observed at the hydrogen evolution potential. In acid solution (50 mM H$_{2}$SO$_{4}$, 0.1 
M HClO$_{4}$), the 0.1 $G_{0}$ peak was observed in the conductance histogram, in addition to the 0.5 $G_{0}$  
peak at the hydrogen evolution potential. Figure~\ref{fig3} shows the typical conductance traces and 
histogram of the Au nano contacts in 50 mM H$_{2}$SO$_{4}$ solution during breaking the contact at the 
hydrogen evolution potential. Reversible transitions of the conductance between 1 $G_{0}$  and 0.5 $G_{0}$, 
0.5 $G_{0}$  and 0.1 $G_{0}$  were observed in the conductance traces. This conductance fluctuation resulted in 
the 0.5 $G_{0}$  and 0.1 $G_{0}$  peaks in the conductance histogram. The 0.1 $G_{0}$  peak 
in Fig.~\ref{fig3}(b) seems very 
sharp compared to other 1 $G_{0}$  and 0.5 $G_{0}$  peaks. The half width at half maximum ($W$) of the 
histogram was 0.14, 0.15, 0.08 $G_{0}$ for the 1, 0.5, 0.1 $G_{0}$  peaks, respectively. On the other hand, the 
relative peak width ($W$ / conductance ratio) was 0.14, 0.28, 0.74 for the 1, 0.5, 0.1 $G_{0}$  peaks, 
showing that the 0.1 $G_{0}$  peak became broader than the 0.5 $G_{0}$  and 1 $G_{0}$  peaks.

\begin{figure}
\begin{center}
\leavevmode\epsfysize=60mm \epsfbox{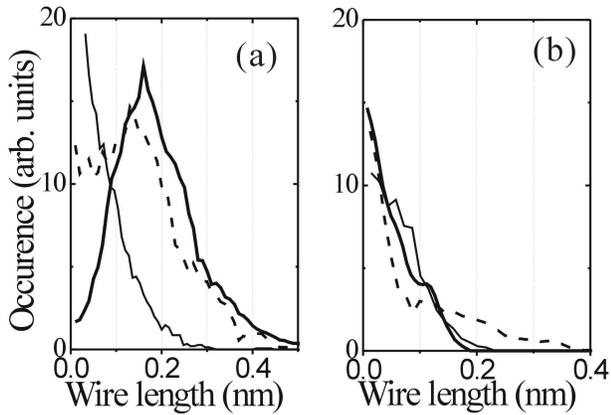}
\caption{
Length histogram of the Au mono atomic contact in 0.1M Na$_{2}$SO$_{4}$ solution. The thin, thick, 
and dotted lines show the result at $\phi$= -0.6 V(double layer potential), $\phi$= +0.8 V (oxide formation 
potential), and $\phi$= -1.0 V (hydrogen evolution potential), respectively. The length was defined as 
the distance between the points at which the conductance dropped below (a) 1.3 $G_{0}$  and 0.05 $G_{0}$, (b) 
1.3 $G_{0}$  and 0.8 $G_{0}$ , respectively.} \label{fig4}
\end{center}
\end{figure}

The electrochemical potential dependence of the stability of the Au atomic contact was 
investigated in 0.1 M Na$_{2}$SO$_{4}$ solution. Figure~\ref{fig4}(a) shows the distribution of lengths for the last 
conductance plateau ($D_{last}$). The length of the last plateau was defined as the distance between the 
points at which the conductance dropped below 1.3 $G_{0}$  and 0.05 $G_{0}$ , respectively. At the double 
layer potential, the contact broke within 0.2 nm. The contact could be stretched a longer distance at 
the hydrogen evolution and oxide formation potential. The average length of the Au mono atomic 
contact was 0.08 nm, 0.15 nm, and 0.18nm at the double layer, hydrogen evolution, and oxide 
formation potentials, respectively. At the hydrogen and oxygen potential, the mono atomic contact 
could be stretched 0.4 nm in length. Considering the Au-Au distance of 0.25 nm in the Au mono 
atomic contact obtained at low temperature in UHV \cite{14}, the Au mono atomic wire could be formed 
in solution at the hydrogen evolution and oxide formation potential. Of course, the long stretched 
length did not directly indicate the formation of a mono atomic wire; the stem part of the Au contact 
might be deformed during the stretching. Although the formation of the Au mono atomic wire was 
not clear, the Au mono atomic contact was found to be stabilized at the hydrogen evolution and 
oxide formation potential.

In order to investigate the transformation process of the Au mono atomic contact in more detail, 
the distribution of lengths was analyzed for the 1 $G_{0}$  plateau, which corresponded to a clean Au 
mono atomic contact \cite{2} ($D_{1G}$: see Fig.~\ref{fig4}(b)). The length of the 1 $G_{0}$  plateau was defined as the 
distance between the points at which the conductance dropped below 1.3 $G_{0}$  and 0.8 $G_{0}$, 
respectively. At the double layer and hydrogen evolution potential, the $D_{1G}$ was similar to the $D_{last}$. 
The 1 $G_{0}$  plateau could be stretched 0.4 nm in length at the hydrogen evolution potential. On the 
other hand, the 1 $G_{0}$  plateau was very short ($<$ 0.2 nm) at the oxide formation potential, although the 
last conductance plateau could be stretched 0.4 nm in length. The $D_{1G}$ at the oxide formation 
potential was close to the $D_{1G}$ and $D_{last}$ at the double layer potential. At the hydrogen evolution 
potential, the close agreement between the $D_{1G}$ and $D_{last}$ indicated that the formation of the structure 
showing 0.5 $G_{0}$  would not relate with the stabilization of the Au mono atomic contact at the 
hydrogen evolution potential. At the oxide formation potential, the conductance of the Au contact 
continuously decreased stepwise at various conductance values after showing 1 $G_{0}$. Combing with 
this continuous change in conductance value, the close agreement between the $D_{1G}$ at the oxide 
formation potential and the $D_{1G}$ at the double layer potential suggested the following transformation 
process of the Au atomic contact. Initially, the clean Au mono atomic contact was formed during 
the stretching the contact. Then, the structure showing conductance value below 1 $G_{0}$  would be 
formed at the oxide formation potential.

Here, the present experimental results are compared with our previously reported results 
measured in 0.1 M Na$_{2}$SO$_{4}$ and 50 mM H$_{2}$SO$_{4}$ solution \cite{10}. 
In our previous study, it was revealed 
that the electrochemical potential significantly affected the stability and conductance of the Au 
mono atomic contact. At the hydrogen evolution potential, the conductance histogram showed the 
0.5 $G_{0}$  peak, whose intensity could be tuned by the electrochemical potential. The stability of the 
Au mono atomic contact could be also tuned by the electrochemical potential. As the potential of 
the Au electrodes was scanned from $\phi$ = 0.5 V to negative, the average length of the last 
conductance plateau decreased at $\phi$ = -0.2 V, and then reached a minimum value at $\phi$ = -0.4 V in 
0.1 M Na$_{2}$SO$_{4}$ solution. Polarization more negative than $\phi$ = -0.6 V led to the recovery of the 
length. At the hydrogen evolution potential ($\phi$ $<$ -0.6 V), the 1 nm long Au mono atomic wire could 
be occasionally fabricated. The distribution of lengths for the last conductance plateau was 
investigated at the hydrogen evolution potential. The same behavior was observed again in the 
present system. In addition to previously reported characteristics, several interesting features 
appeared in the present system. First, the Au atomic contact showed the third conductance value 
(various conductance values below 1 $G_{0}$) at the oxide formation potential. Second, the Au mono 
atomic contact showed not only 0.5 $G_{0}$  but also 0.1 $G_{0}$  
in acid solution at the hydrogen evolution 
potential. This 0.1 $G_{0}$  peak was not apparent in the conductance histogram in Fig. 6 of Reference 
\cite{10}, because the intensity of the background around 0.1 $G_{0}$  
was beyond the vertical axis range of 
the conductance histogram in Fig. 6 of Reference \cite{10}. Third, the Au atomic contact was stabilized 
at the oxide formation potential. Fourth, the distribution of lengths for the last conductance plateau 
was investigated at the double layer and oxide formation potential. The distribution of lengths was 
also shown for the 1 $G_{0}$  plateau, which provided information about the transformation process of the 
Au mono atomic contact during the stretching the contact. Fifth, the distribution of lengths for the 
conductance plateau was precisely determined by the statistical analysis with a large number of 
measurements for 10 different samples (20000 traces). In our previous report, the distribution of 
lengths for the last conductance plateau was obtained from limited number of conductance traces 
($\sim$1000 traces) for one extraordinarily stable sample at the hydrogen evolution potential \cite{10}. 
Although very stable Au mono atomic wire occasionally could be fabricated, most of the Au atomic 
contact broke within 0.4 nm in length. So, the obtained distribution of lengths for the last 
conductance plateau shifted to the shorter distance compared to the previous result. The effect of 
hydrogen and oxygen on the Au mono atomic contacts is discussed based on these new findings and 
improvements.

The conductance behavior of the Au mono atomic contact is compared with that in UHV. The 
effect of hydrogen and oxygen on the Au mono atomic contact was investigated at low temperature 
in UHV \cite{4,5}. The broad feature appeared below 1 $G_{0}$  in the conductance histogram of the Au 
contacts after the introduction of hydrogen gas. The feature below 1 $G_{0}$  was much smaller than the 1 
$G_{0}$  peak \cite{4}. The conductance histogram did not change 
with the introduction of oxygen gas \cite{5}. On 
the other hand, in solution, clear 0.5 $G_{0}$  and 0.1 $G_{0}$  peaks appeared in the conductance histogram at 
the hydrogen evolution potential. A broad feature appeared below 1 $G_{0}$  at the oxide formation 
potential. These results suggest that a specific structure of the Au mono atomic contact with a 
well-defined fractional conductance value (0.5 $G_{0}$  or 0.1 $G_{0}$) was formed at the hydrogen evolution 
potential, and various structures in which oxygen strongly interacted with the atomic contact were 
formed at the oxide formation potential. The electrochemical potential determines the potential 
energy of electrons of the metal nano contact, resulting in the control of the bonding strength 
between the metal atoms, and of the interaction of the metals with molecules in the surrounding 
medium. These facts make it possible to use the environment to set the metal contacts, which cannot 
be set in other environments such as in UHV and in air, leading to successful fabrication of very 
stable metal nano structures showing the conductance quantization which cannot be observed in 
UHV. 

The structure of the Au mono atomic contact at the hydrogen evolution and oxide formation 
potential is discussed based on the previously reported experimental result of a flat Au surface and 
theoretical calculation result. The hydrogen and oxygen evolution reaction on the flat Au electrodes 
in solution has been investigated by the analysis of current-potential curves \cite{11,12,13}. When the 
electrochemical potential of the Au electrode is kept positive ($\phi$ $>$ 1.6 V) in acidic solution, oxygen 
evolution has been proposed to proceed through the following process. First, hydroxyl ions adsorb 
onto the Au surface forming surface hydroxides (1), second, surface hydroxides convert to oxides 
(2), and third, O$_{2}$  (gas) desorbs from the surface (3).

Au+H$_{2}$O $\rightarrow$ Au-OH + H$^{+}$ + e$^{-}$ (1)

2Au-OH $\rightarrow$ Au-O + Au + H$_{2}$O  (2)  

2Au-O $\rightarrow$ 2Au+O$_{2}$   (3)

When the electrochemical potential of the Au electrode is kept negative ($\phi$ $<$ -0.3V) in acid solution, 
hydrogen evolution proceeds through the following process. First, protons adsorb onto the Au 
surface (4), and H$_{2}$  (gas) desorbs via surface diffusion and recombination of two adsorbed H atoms 
(5) or a combination of adsorbed H atoms and proton (5').

Au+H$_{3}$O$^{+}$ + e$^{-}$ $\rightarrow$ Au-H +H$_{2}$O  (4)

2Au-H $\rightarrow$ 2Au+H$_{2}$  (5)

Au-H+ H$_{3}$O$^{+}$ + e$^{-}$ $\rightarrow$ Au + H$_{2}$ + H$_{2}$O   (5')

In the CV of the Au electrode in 0.1 M Na$_{2}$SO$_{4}$ solution (Fig.~\ref{fig1}(b)), the main anodic peak above $\phi$= 
0.8 V and an increase in the oxidation current above $\phi$= 1.3 V correspond to the oxide formation 
process (2) and the O$_{2}$  (gas) desorption process (3). 
As for the surface oxide, higher oxide (Au$_{2}$O$_{3}$) 
or hydroxide (Au(OH)$_{3}$) was proposed \cite{13}. The increase in redox current below $\phi$= -0.7 V 
corresponds to the H$_{2}$  (gas) desorption process (5). The surface coverage of hydrogen was estimated 
to be very small ($<$ 0.3 $\%$)\cite{12}. Here, it should be noticed that the detail reaction mechanism of the 
oxygen and hydrogen evolution reaction on the Au electrode has not been fully understood up to 
now. In addition, the reaction on the Au mono atomic contact might not be the same as that on the 
flat Au electrode. However, the observed conductance behavior and the reaction on the flat metal 
surface strongly suggested that the oxygen or hydrogen is adsorbed on or incorporated into the Au 
mono atomic contact at the oxide formation and hydrogen evolution potential.

The structure of the Au mono atomic contact is discussed with the previously reported 
theoretical calculation result. The interaction between the Au mono atomic wire and oxygen or 
hydrogen has been investigated with theoretical calculations \cite{15,16,17}. Theoretical calculation results 
showed that the hydrogen 1s orbital or oxygen 2p orbital effectively hybridized with the Au 5d and 
Au 6s orbitals. Due to the strong interaction, oxygen and hydrogen molecules dissociated on the Au 
mono atomic wire, and then were stably incorporated into the Au mono atomic wire. In the 
hydrogen (oxygen) incorporated wire, we can expect electron transfer from hydrogen to Au (from 
Au to oxygen). The Au wire would be, thus, positively (negatively) charged for the hydrogen 
(oxygen) incorporated wire. Because of the formation of hydrogen or oxygen incorporation into the 
wire, the conductance of the Au mono atomic wire decreased to 0.6-0.01 $G_{0}$, possibly due to 
scattering or interference of conducting electrons in the wire. No preferential atomic configurations 
were found for the hydrogen or oxygen incorporated Au mono atomic wire. Jelinek et al. calculated 
the conductance of the Au mono atomic wire in which a hydrogen atom or an undissociated 
molecule adsorbed on it \cite{16}. 
The conductance value was calculated to be 0.7-0.5 $G_{0}$  and 1.05-0.95 
$G_{0}$  for the Au atomic contact in which a hydrogen atom and molecule adsorbed on it, respectively. 
The adsorbed atomic hydrogen effectively affected the conductance of the Au mono atomic wire, as 
is the case for the incorporation of hydrogen into the Au wire. On the other hand, the conductance 
value of the Au atomic wire with an adsorbed undissociated hydrogen molecule did not change 
from that of the clean Au atomic wire (1 $G_{0}$). The adsorbed undissociated hydrogen molecule would 
not affect the conductance of the Au mono atomic wire, in contrast with the adsorbed hydrogen 
atom and incorporated hydrogen.

Now, we try to propose the structure model of the Au mono atomic contact at the hydrogen 
evolution and oxide formation potential based on the above discussion. At the oxide formation 
potential, various conductance values were observed for the Au mono atomic contact 
(see Fig.~\ref{fig2}(c,f)). This conductance behavior agreed with the calculated conductance behavior of the oxygen 
incorporated Au wire \cite{15,16,17}. So, the conductance value below 1 $G_{0}$  would originate from the 
oxygen incorporated Au mono atomic contact. Combing with the result of the plateau length 
analysis showing the long last conductance plateau and short 1 $G_{0}$  plateau which was comparable to 
the 1 $G_{0}$  plateau at the double layer potential, the following transformation process of the Au atomic 
contact could be proposed. Initially, the clean Au mono atomic contact was formed during the 
stretching the contact. Then, the oxygen incorporated Au mono atomic contact would be formed. In 
contrast to the oxide formation potential, reversible transition between 1 $G_{0}$  and the well defined 
fractional conductance value (0.5 $G_{0}$) were observed in the conductance trace at the hydrogen 
evolution potential (see Fig.~\ref{fig2}(a)). The reversible transition suggests the successive adsorption and 
desorption of hydrogen on the Au mono atomic contact. In the case of the atom or molecular 
adsorption on flat metal surfaces, atoms or molecules often adsorb with a fixed atomic 
configuration \cite{17}. Therefore, hydrogen could also adsorb on the Au mono atomic contact with a 
certain fixed atomic configuration showing a fixed conductance value. When the hydrogen 
incorporates into the Au contact, the Au mono atomic contact would not show a fixed conductance 
value, in contrast to the present experimental result. So, the hydrogen adsorbed Au mono atomic 
contact would be formed at the hydrogen evolution potential. As discussed in the previous section, 
the surface coverage of hydrogen is very low on the Au electrode \cite{12}, and thus, the number of 
hydrogen atoms or molecules adsorbed on the Au mono atomic contact would be very small. The 
conductance value was calculated to be about 1.0 $G_{0}$  and 0.6 $G_{0}$  for the Au atomic contact in which 
one hydrogen molecule and atom adsorbed on it, respectively \cite{16}. Therefore, 0.5 $G_{0}$  would 
originate from the Au mono atomic contact, in which one hydrogen atom adsorbed on it.

The above discussion about the Au mono atomic contact with an adsorbed hydrogen atom 
could be supported by the conductance behavior of the Au nano contacts in acid solution. In acid 
solution, the rate determining process of the hydrogen evolution reaction on Au electrodes is the 
surface diffusion and desorption process of H$_{2}$  gas (5). The surface coverage of hydrogen atoms on 
the Au surface in acid solution is expected to be higher than that in neutral solution 
(Na$_{2}$SO$_{4}$) \cite{11}. 
Therefore, more than one hydrogen atom occasionally could adsorb on the Au mono atomic contact 
in acid solution under the hydrogen evolution reaction. The Au mono atomic contact with two 
adsorbed hydrogen atoms would show a smaller conductance value than that with one hydrogen 
atom. In 50mM H$_{2}$SO$_{4}$ and 0.1 M HClO$_{4}$, the 0.1 $G_{0}$  
peak appeared in the conductance histogram in 
addition to the 0.5 $G_{0}$  peak at the hydrogen evolution potential 
(see Fig.~\ref{fig3}(b)). The 0.1 $G_{0}$  peak 
would originate from the Au mono atomic contact in which two atomic hydrogen atoms adsorbed 
on it. This hypothesis about the origin of 0.1 $G_{0}$  peak can be supported by the conductance behavior 
of the Au atomic contact in acid solution at the hydrogen evolution potential. In the conductance 
traces, the reversible transitions of the conductance between 0.5 $G_{0}$  and 0.1 $G_{0}$, 
1 $G_{0}$  and 0.5 $G_{0}$  
were observed (see Fig.~\ref{fig3}(a)). The reversible transitions of the conductance between 0.5 $G_{0}$  and 0.1 
$G_{0}$  suggested the successive adsorption and desorption of hydrogen atom on the Au mono atomic 
contact, as is the case for the reversible transitions of the conductance between 1 $G_{0}$  and 0.5 $G_{0}$. The 
appearance of the 0.1 $G_{0}$  peak in the conductance histogram also suggested that the origin of the 
fractional conductance value was not the dimerized wire proposed in previous study \cite{10}. It is 
because the dimerized wire would show one fixed conductance value ($\sim$0.5 $G_{0}$ ). But, the Au mono 
atomic contact showed both 0.1 $G_{0}$  and 0.5 $G_{0}$  in acid solution under the hydrogen evolution.

Improved stability of the Au mono atomic contact at the hydrogen evolution and oxide 
formation potential is discussed. In UHV at 4 K, stabilization of Au or Ag mono atomic wire was 
observed by oxygen incorporation into the wire. As for the Ag, a 2nm long mono atomic wire could 
be formed by oxygen incorporation into the wire, while a clean Ag forms only short atomic contact 
\cite{5}. The theoretical calculation results showed that the incorporated oxygen strengthened Au-Au 
and Ag-Ag bonds in the wire, leading to stabilization of the Au and Ag mono atomic contact \cite{18}. 
The stabilization of the metal nano wires by molecular adsorption was also observed for Au, Fe, Co, 
and Ni nano wires \cite{19}. This stabilization was explained by the decrease in the surface energy 
caused by the molecular adsorption on the metal nano wires. Therefore, the stabilization of the Au 
mono atomic contact at the oxygen formation potential could be explained by the increase in the 
bond strength in the atomic contact caused by the oxygen incorporation into the contact. The 
stabilization of the contact at the hydrogen evolution potential could be explained by the decrease in 
the surface energy caused by the hydrogen adsorption on the Au mono atomic contact. The plateau 
length analysis revealed that the 1 $G_{0}$  plateau was much longer at the hydrogen evolution potential 
than the double layer potential. Since the conductance of the Au mono atomic contact with an 
adsorbed hydrogen molecule was calculate to be about 1 $G_{0}$  \cite{16}, the long 1 $G_{0}$  plateau at the 
hydrogen evolution potential suggested that the Au mono atomic contact would be stabilized by the 
adsorption of a hydrogen molecule on the Au mono atomic contact.

In the present study, the structure models of the Au mono atomic contact at the hydrogen 
evolution and oxide formation potential were proposed based on the experimental and theoretical 
calculation result. Although our model of the hydrogen adsorbed and oxygen incorporated Au mono 
atomic contact could explain the experimental results, we did not obtain direct evidence for the 
formation of the hydrogen adsorbed and oxygen incorporated Au mono atomic contact proposed in 
this study. In addition, it is not completely clear whether a hydrogen (or oxygen) molecule or atom 
adsorbed on (or incorporated into) the Au mono atomic contact. The model proposed in this study is 
one of the possible models. Further investigation is needed to fix the structure formed at the 
hydrogen evolution and oxide formation potential. Although the structure model is not defined yet, 
our experimental results clearly showed the existence of hydrogen or oxygen at the Au mono atomic 
contact as a change of conductance. Up to now, there is little experimental results which directly 
show the existence of hydrogen on the Au surface under the hydrogen evolution reaction. Based on 
the qualitative analysis of current-potential curves, the surface coverage of hydrogen has been 
estimated. Thus, our present study might shed light on the understanding the mechanism of the 
hydrogen evolution reaction on the Au surface, as well as the conductance modulation of metal 
nanowire.

\section{CONCLUSIONS}
\label{sec4}

The conductance behavior of the Au mono atomic contact was studied under the 
electrochemical potential control. The stability and conductance of the Au contact could be fully 
controlled by the electrochemical potential. The conductance could be defined among three 
different conductance states at respective potentials: 1 $G_{0}$  at the double layer potential, 0.5 $G_{0}$  at 
hydrogen evolution and conductance below 1 $G_{0}$  at the oxide formation potential. Based on the 
comparison between the conductance behavior and previously documented surface processes, we 
proposed the structural model. An oxygen would incorporate into the contact at the oxide formation 
potential. Atomic hydrogen would adsorb on the Au mono atomic contact at the hydrogen evolution 
potential. At the hydrogen evolution and oxide formation potential, the Au mono atomic contact was 
stabilized. The stabilization could be explained by the oxygen incorporation into the contact or 
hydrogen adsorption on the contact. Control of the three conductance states, stabilization of a fixed 
conductance state with hydrogen, and the appearance of the structure showing strong interaction 
between Au or oxygen were observed only for the metal contact under the electrochemical potential 
control, which could not be observed in other environments. The present results showed that 
electrochemical potential is one of the promising external fields for controlling the properties of 
nano scale materials.

\acknowledgments{
This work was partially supported by a Grant-in-Aid for Scientific Research A (No. 16205026) 
and Grant-in-Aid for Scientific Research on Priority Areas (No. 17069001) and Global COE 
Program (Project No. B01: Catalysis as the Basis for Innovation in Materials Science) from MEXT, 
and Asahi Glass Foundation.}

\end{document}